\documentclass[prl,twocolumn,aps,amssymb,footinbib]{revtex4}
\usepackage{amssymb}

\usepackage{amssymb}

\usepackage{graphicx}
\usepackage{amsmath}
\usepackage{times}

\begin{document}

\title{Pseudo-fermionization  of 1-D bosons in optical lattices
}
\author{  Guido Pupillo, Ana Maria Rey, Carl J. Williams and Charles W. Clark}
\affiliation{National Institute of Standards and Technology,
Gaithersburg, MD 20899}
\date{\today}

\begin{abstract}

We present a model that generalizes the Bose-Fermi mapping for
strongly correlated 1D bosons in an optical lattice, to cases in
which the average number of atoms per site is larger than one. This
model gives an accurate account of equilibrium properties of such
systems, in parameter regimes relevant to current experiments. The
application of this model to non-equilibrium phenomena is explored
by a study of the dynamics of an atom cloud subject to a sudden
displacement of the confining potential. Good agreement is found
with results of recent experiments. The simplicity and intuitive
appeal of this model make it attractive as a general tool for
understanding bosonic systems in the strongly correlated regime.

\end{abstract}
\maketitle {\it Introduction} Cold bosonic atoms in optical lattices
have recently been used to create quasi-one dimensional systems
\cite{Tolra,Moritz,Soferle,Weiss,Paredes,trey}. In such experiments,
arrays of one dimensional tubes are realized by first magnetically
trapping a Bose-Einstein condensate (BEC) in a parabolic potential,
and then imposing upon it a deep 2D optical lattice, which restricts
atomic motions to 1D. These defect-free highly controllable  atomic
systems offer an excellent opportunity to directly study strongly
correlated regimes.

For low densities or large interaction strengths, a 1D gas of
ultracold bosons behaves as a gas of impenetrable particles, known
as a Tonks-Girardeau (TG) gas. Ref. \cite{Girardeau} shows there
is a one to one mapping between the eigenenergies and
eigenfuntions of
 TG bosons  and the ones of non-interacting fermions, known as fermionization. Two recent experiments
\cite{Weiss,Paredes} successfully reached this parameter regime. In
Ref.\cite{Paredes} the  TG regime was achieved by  adding an optical
lattice in the tubes' direction which  increases the effective mass
and therefore the ratio between interaction and kinetic energy. When
the lattice is present and for large enough interactions  a
commensurate homogenous  system not only fermionizes  but also
undergoes the superfluid to Mott insulator (MI) transition
\cite{Fisher}. In the presence of a parabolic trap the MI can be
realized for any number of particles \cite{Ana2005,Rigol}.

In the TG regime where there is at most one particle per site,
single-particle solutions of the periodic plus parabolic potential
have been successfully used to describe equilibrium and
non-equilibrium properties of the system, both in the presence and
in the absence of the MI \cite{Ana2005,Rigol}. However, 1D
experiments have been realized in a parameter regime where on-site
particle densities are larger than one and standard fermionization
is inapplicable \cite{Soferle,trey}. The study of these regimes
beyond mean-field has relied mainly on numerical simulations, such
as quantum Monte-Carlo or density matrix renormalization group
techniques \cite{Batrouni,Prokofev,Kollath,Daley}. Here we show that
even when the on-site density is larger than one, for a wide range
of conditions single-particle solutions can still describe strongly
correlated regimes. We call this single-particle approach {\it
pseudo-fermionization} (PF). Comparison with exact Monte-Carlo
simulations shows that in the appropriate parameter regimes PF can
be used to accurately reproduce equilibrium properties such as the
density profile, the momentum distribution, and the ground-state
energy.  In the final section we extend our model to treat the
non-equilibrium dipole oscillations of atoms subject to a displaced
potential, as has been realized in recent experiments
\cite{Soferle,trey}. The accuracy and simplicity of the PF method
suggest that it is a useful
tool for understanding strongly correlated bosonic systems.\\

{\it Bose-Hubbard Hamiltonian} The Bose-Hubbard ({\it BH})
Hamiltonian describes the system's dynamics when the lattice is
loaded such that only the lowest vibrational level of each lattice
site is occupied and tunneling occurs only between nearest-neighbors
\cite{Jaksch}
\begin{equation}
H= -J\sum_{\langle i,j\rangle
}\hat{a}_i^{\dagger}\hat{a}_{j}+\frac{U}{2}\sum_{j}\hat{n}_j(\hat{n}_j-1)
+\Omega \sum_{j} j^2 \hat{n}_j
.\\
\label{EQNBHH}
\end{equation}
Here $\hat{a}_j$ is the bosonic annihilation operator of a particle
at site $j$, $\hat{n}_j=\hat{a}_j^{\dagger}\hat{a}_{j}$, and the sum
$\langle i,j\rangle$ is over nearest neighbors. In Eq.(\ref{EQNBHH})
the hopping parameter $J$, and the on-site interaction energy  $U$
are functions of the lattice depth $V_o$. $\Omega$ is proportional
to the curvature of the parabolic potential.

{\it Homogeneous system} For a homogeneous system ($\Omega=0$) with
$N$ atoms and $M$ sites the spectrum is fully characterized by the
ratio between interaction and hopping energies $\gamma=U/J$, and by
the filling factor $N/M=(n-1)+\Delta N/M$. Here $n$ is the smallest
integer larger than $N/M$ and $\Delta N<M$. If the lattice is
commensurately filled  ($\Delta N=0$) and $\gamma > \gamma_c (n-1)
$, the ground state is a Mott insulator with reduced number
fluctuations. The critical value  $\gamma_c $ is about 4 according
to numerical results \cite{Monien}. For the incommensurate case,
{\it i.e.} $\Delta N>0$, the ground state is a superfluid. In this
case, if $\gamma/n \gg \gamma_c $ the extra atoms can be thought of
as TG bosons with effective hopping energy $nJ$ on top of a Mott
state with filling factor $n-1$. This is justified because the
population of states with more than $n$ atoms per site is suppressed
by a factor of the order of $J/U$. The lowest $M!/((\Delta
N)!(M-\Delta N)!)$ eigenstates and eigenenergies can then be
described by means of the standard Bose-Fermi mapping with $J$
replaced by $n J$. We refer to this approach as {\it
pseudo-fermionization}.

{\it Inhomogeneous system} When the parabolic trap is present
($\Omega > 0$), the density profile of the atomic cloud is
determined by an interplay of $U$, $J$, $\Omega$ and $N$. The system
is fermionized if \cite{Ana2005}
\begin{equation}
\gamma > \gamma_c, \quad U > \Omega((N-1)/2)^2.\label{ferm}
\end{equation}

While the first inequality is the same as for homogeneous lattices,
the second one is specific for trapped systems and is necessary to
suppress multiple occupancy of single sites. In this fermionized
regime, for decreasing $J$ the density at the trap center increases,
and, when the condition $2 J\lesssim \Omega((N-1)/2)^2 $ is
satisfied, sites around the trap center  begin to have unit filling
\cite{Rigol,Ana2005}.
 If the inequality
\begin{equation}
J < \Omega N,\label{mot}
\end{equation}
is also satisfied, the ground state is a unit-filled Mott state in
all $N$ sites. In this case, fluctuations occur mainly at the edge
of the density distribution, due to tunneling of atoms to empty
sites. Such fluctuations are proportional to $\Omega N$, which is
the trap gradient at the site $(N-1)/2$. Residual fluctuations at
the trap center are due to mixing of particle-hole excitations and
are proportional to $2 \sqrt{2} (J/U)$.

For $\Omega((N-1)/2)^2 \gtrsim U$  it is energetically favorable for
atoms to pile up at the trap center. In fact, superfluid and Mott
insulator phases with different filling factors can coexist, due to
the interplay between on-site interactions and the external
potential \cite{Jaksch,Batrouni, DeMarco}. In the trivial case
$J=0$, Fock states with a definite number of atoms in each site are
eigenstates of the Hamiltonian. The density profile results into a
``cake''-like structure with maximal occupation $n_{max}$ at the
trap center if

\begin{equation}
 2 H^{(1/2)}_{n_{max}-1}<\left[(N-n_{max})
\sqrt{\Omega/U} \right] <2 H^{(1/2)}_{n_{max}},\label{harmonic}
\end{equation}

\noindent where $H^{(1/2)}_{n}\equiv\sum_{i=0}^{n} \sqrt{i}$
\cite{DeMarco}. The number of atoms $N_n$ in the $n^{th}$ "cake"
layer is given by $(N_{1}-N_n)(N_{1}+N_n-2)=4U (n-1)/\Omega$ with
the normalization condition $\sum_{n=1}^{n_{max}} N_n=N$. Similar to
the PF in the homogeneous case, we can view the atoms in each layer
as creating an independent Mott state with unit filling. For finite
but small values of $J$, all layers except the top one can still be
thought of as an independent Mott state with filling factor one if
\begin{equation}
\gamma > \gamma_c (n_{max}-1), \quad \Omega N_{n_{max-1}}> J
(n_{max}-1).\label{pseudo}
\end{equation}
\noindent In analogy to Eq. (\ref{ferm}) the first inequality
insures that in each layer with $n<n_{max}$ the average kinetic
energy required for one atom to hop from one site to the next is
insufficient to overcome the potential energy cost and therefore
particle-hole excitations are suppressed by a factor of order $n J /
U$. The second inequality guarantees that in all but the
$n_{max}^{th}$ layer number fluctuations are confined mainly at the
edge of the Mott state, in analogy to  Eq.(\ref{mot}).

Under a wide range of relevant experimental parameter regimes the
above conditions are satisfied and fluctuations in the various
layers do not spatially overlap. If also
\begin{equation}
\gamma > \gamma n_{max},\label{pseu}
\end{equation}
\noindent to a very good approximation atoms in all layers can be
treated as TG bosons with an effective hopping energy $n J$.
 Notice that atoms in the top layer do not need to be localized.
Under these conditions, single-particle solutions can be
successfully used to obtain expressions for all many-body
observables. This is the generalization of PF to trapped systems.

In the following we show the success and the limitations of this
approach, by applying it to a system with the same parameters as
the ones used in an experiment recently performed at NIST
\cite{trey}.

{\it{Numerical comparisons}} In the NIST experiment \cite{trey}
approximately $N_T=1.4 \times 10^5$ $^{87}$Rb atoms were trapped in
an array of one-dimensional tubes with $N \simeq 80$ atoms in the
central tube. An additional periodic potential was added along the
direction of the tubes and  its  depth $V_o$  was varied for
different experiments. The frequency of the parabolic potential in
the tubes' direction was such that  $\Omega = 7.4 \times 10^{-4}
E_R$, with $E_R$  the photon recoil energy. Here we focus on the
central tube only and consider $V_o> 2 E_R$, where the tight-binding
Hamiltonian Eq.(\ref{EQNBHH}) is expected to be valid. For these
parameters, Eq. (\ref{harmonic}) yields $n_{max}=2$.

\begin{figure}[t]
\begin{center}
\leavevmode {\includegraphics[width=3. in]{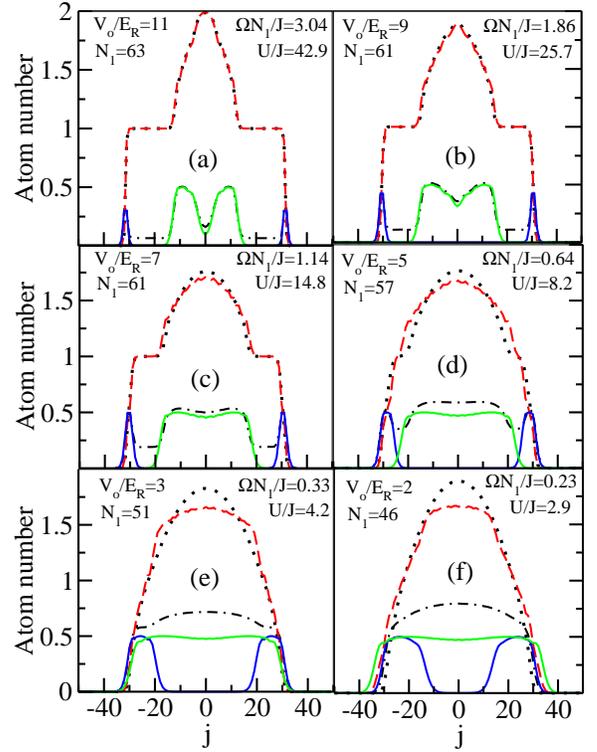}}
\end{center}
\caption{
 Local densities $\langle\hat{\rho}_j\rangle$ and fluctuations $\langle \Delta
\hat{\rho}^{(n)}_j\rangle$ as a function of the site index $j$.
Dotted(black) and dashed(red) lines are the numerical and analytical
densities, respectively. The dashed-dotted(black) line is the
numerical fluctuation, while solid-black(blue) and solid-grey(green)
lines are the analytical fluctuations for the lower($n=1$)
and upper($n=2$) layers, respectively. 
}\label{dens}
\end{figure}

In Fig.~\ref{dens} we show comparisons between the density and
number fluctuations for the central tube  calculated by using the PF
approximation and exact quantum Monte-Carlo numerical simulations
based on the Worm algorithm \cite{Proko}, for lattice depths
$V_o/E_R=11(a)$,$9(b)$,$7(c)$,$5(d)$,$3(e)$,$2(f)$. In the numerical
simulations the temperature is $0.01 J$. In the plots the
dotted(black) and dash-dotted(black) lines correspond to the density
and number fluctuations as numerically calculated by using the
Monte-Carlo code, respectively. The dashed(red), the
solid-black(blue) and solid-grey(green) lines correspond to the
density $\langle\hat{\rho}_j\rangle$, and to the number fluctuations
for the atoms in the first $\langle \Delta \hat{\rho}^{(1)}_j
\rangle$ and second layer $\langle \Delta
\hat{\rho}^{(2)}_j\rangle$, as calculated with the PF model
respectively. In particular, the density is given by
$\langle\hat{\rho}_j\rangle=\sum_{s=0}^{N_1-1}|{f_j^1}^{(s)}|^2+\sum_{s=0}^{N_2-1}|{f_j^2}^{(s)}|^2$
where $\{{f_j^{n=1,2}}^{(s)}\}$  are the
 $s^{th}$ single-particle eigenmodes
of Eq.(\ref{EQNBHH})  with hopping energies $nJ$ respectively and
$j$ is the lattice site index. The fluctuations are given by
$\langle \Delta
\hat{\rho}^{(n)}_j\rangle=\sqrt{\sum_{s=0}^{N_n-1}|{f_j^n}^{(s)}|^2-\left(\sum_{s=0}^{N_n-1}|{f_j^n}^{(s)}|^2\right)^2}$.

The conditions for PF to be applicable, Eqs.(\ref{pseudo}) and
(\ref{pseu}), are strictly valid for $V_o \gtrsim 5 E_R$.
Consistently, Fig.~\ref{dens}$(a-c)$ shows that for $V_o/E_R
=11,9,7$ the density profile and number fluctuations are very well
reproduced by the model, except for the finite value of the
fluctuations in the flat region of the density profile. These small
fluctuations of order $2 \sqrt{2} J/U$ are due to the particle-hole
excitations which are neglected in the model. The model predicts
that some sites at the trap center have exactly a filling factor of
two when $ 2 (2J) \lesssim \Omega((N_2-1)/2)^2 $. This condition is
fulfilled for $V_o=11 E_R$, and in fact a flat density distribution
with two atoms per site is observed  in Fig.1-(a) at the trap
center, both in the analytical and numerical results. This confirms
the validity of the idea of thinking of the atoms in the second
layer as TG-bosons with effective hopping energy $2 J$.

For $V_o=5 E_R$,$\gamma$ is barely $2 \gamma_c $, and $\Omega
 N_1 < J$ so that fluctuations at the edge of the first
layer extend far enough to overlap  with the ones of the second
layer. We only expect PF to give qualitative predictions in this
regime. For $V_o=3,2 E_R$ the conditions Eqs.(\ref{pseudo}) and
(\ref{pseu}) are not satisfied, and the model fails to reproduce the
exact results. Nevertheless, we notice that the PF model predicts
the formation of a Mott state in the lowest layer for $V_o\gtrsim
3E_R$ because $\gamma \thickapprox \gamma_c$ and $\Omega
(N_1-1)^2/4> 2 J$ at $3E_R$. The numerically obtained fluctuations
show the appearance of a flat region at the cloud's edge for $V_o=
3E_R$. Such flat region signals the formation of a Mott state as  it
evolves for deeper lattices  into the observed dip in the
fluctuations and disappears for shallower lattices
(Fig~\ref{dens}-(e)).

In Fig.2 we compare the  momentum distribution $\rho(k)$ for the
many-body system, solid(black) line, with  the one predicted by the
PF model, dashed(red) line, for $V_o=11(a)$,$9(b)$,$7(c)$, and
$5(d)$. The model curves are given by $\rho(k)=\rho(k)^{(1)}+2
\rho(k)^{(2)}-N_2/M$, where $\rho(k)^{(n=1,2)}$ are the momentum
distributions for $N_n$ TG bosons with effective hopping rate $nJ$,
calculated numerically with the Monte-Carlo code. For all curves,
the height of the central peak is larger and the width at half
maximum is smaller for the exact than for the model solutions. This
is expected because the model neglects correlations between atoms in
the first and second layers. However, the agreement is at least
qualitative for all displayed lattice depths where PF applies. On
the other hand the agreement for the deepest lattice in
consideration is worse than the one found for the local observables
$\langle\hat{\rho}_j\rangle$ and $\langle \Delta
\hat{\rho}_j\rangle$, (Fig 1). This is consistent with previous
observations for standard fermionization \cite{Pollet}.

\begin{figure}[t]
\begin{center}
\leavevmode {\includegraphics[width=2.75 in]{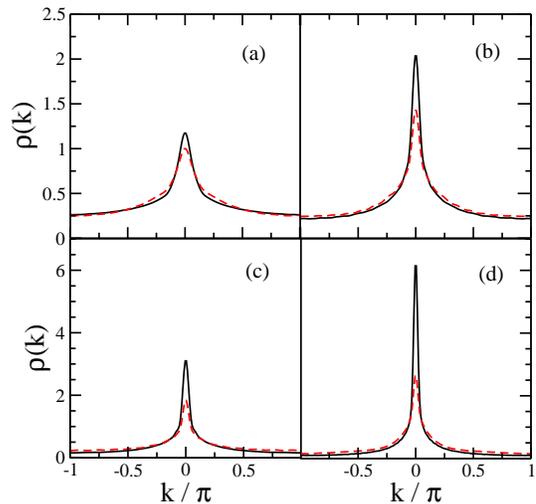}}
\end{center}
\caption{ Momentum profiles $\rho(k)$ as a function of the momentum
$k$, for lattice depths $V_o/E_R=11(a)$, $9(b)$, $7(c)$, and $5(d)$.
The solid(black) and dashed(red) lines are the exact momentum
distibutions and the model momentum distributions, respectively.
 }\label{quasi}
\end{figure}

\begin{figure}[b]
\begin{center}
\leavevmode {\includegraphics[width=3. in]{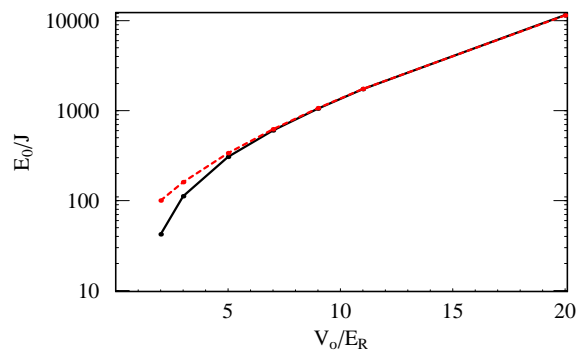}}

\end{center}
\caption{
 Energy as function of $V_o/E_R$. The solid(black)
and dashed(red) lines are the numerical and analytical energies,
respectively.
 }\label{ener}
\end{figure}

In Fig. \ref{ener}  ground state energies are compared as a function
of lattice depth. The  solid(black) line is  calculated numerically
using the Monte-Carlo algorithm and the dashed(red) line using the
PF model  $E=\sum_{s=0}^{N1-1}E_1^{(s)}+
\sum_{s=0}^{N2-1}(U+E_2^{(s)})$. Here, $E_{1,2}^{(s)}$
 are  the $s^{th}$ single-particle eigenenergies of systems  with hopping energies
$J$ and $2J$, respectively. At $V_o=5 E_R$ the model predicts a
ground-state energy which  is $10\%$ larger than the numerical
solution while at $V_o=11 E_R$ the difference decreases to  only
$0.4\%$. This also confirms the validity of the PF model as the
lattice deepens.

{\it{Center of mass oscillations}} In the NIST experiment
\cite{trey}, center of mass oscillations were induced by a sudden
displacement of the harmonic potential by $\delta=8$ lattice sites.
An overdamped motion was observed for lattice depths $V_o \gtrsim 3
E_R$. The damping rate $b$ was obtained by fitting to the formula
$m^*\ddot{x}=-b\dot{x}-m \omega_T^2x$, where $m$ and $m^*$  are the
atomic and effective masses and $\omega_T$ is the magnetic trapping
frequency. Previous theoretical analysis of the damping did not use
real experimental parameters
\cite{Ana2005,polkovnikov1,polkovnikov3,Rigol3} or were not
applicable in the strongly correlated regime
\cite{julio,ruostekoski}. Here we show that the PF approach
reproduces well the experimental results in the overdamped regime.
In Fig.4 the experimental data (black squares) are compared to the
predictions of the PF model. In the model, the center of mass
position of the atoms in the central tube (red dots) is given by
$x(N)=a/N \left[ \sum_{s=0}^{N_1-1} x_1^{(s)}(t,N)+
\sum_{s=0}^{N_2-1} x_2^{(s)}(t,N) \right]$, where
\begin{figure}[t]
\begin{center}
\leavevmode {\includegraphics[width=3.5 in]{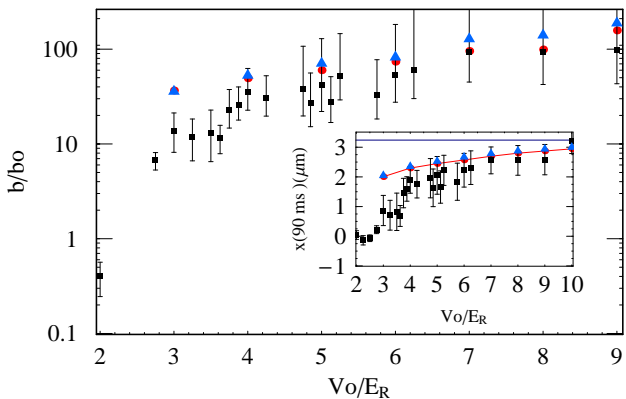}}
\end{center}
\caption{
 Damping rate of dipole oscillations $b$ as a function
of $V_o/E_R$. $b$ is in units of $b_o=2 m \omega_T$. The squares
are the NIST experimental results. The circles(red) are the
model results for the central tube, and the triangles(blue) are the
model results averaged over all the tubes. Inset: center of mass
position after 90 ms.
 }\label{dyna}
\end{figure}

\begin{equation}
x_n^{(s)}(t,N)= \sum_{k,l,j}j {c_l^n} ^{(s)} {c_k^n} ^{(s)} e^{-i
(E_n^{(k)}-E_n^{(l)}) t/\hbar}
{f_j^n}^{(k)}{f_j^n}^{(l)}\label{COM}.
\end{equation}
\noindent Here, $N$ is the number of atoms in the central tube, $a$
is the lattice spacing and the coefficients $ {c_k ^n}^{(s)}
=\sum_j{f_{j}^n}^{(k)} {f_{j-\delta}^n}^{(s)}$ are the projection of
the $s^{th}$ excited state of the displaced potential  onto the
$k^{th}$ excited  eigenstate of the undisplaced potential for atoms
in the $n=1,2$ layer, respectively. The PF model is expected to give
an accurate description of the center of mass oscillations if during
the dynamics the atoms in the first and second layer can still be
treated as independent objects, {\it i.e.} when number fluctuations
in the two layers do not overlap during the evolution. Because in
the experiment the measured damping rate was an average over all the
tubes, in Fig.4 we also plot the model's prediction for the average
damping rate (blue triangles). The latter was calculated assuming
that all tubes evolve independently and therefore that $x(t)=\left[
\sum_{\mathcal{N}=1}^{N} \mathcal{N} P(\mathcal{N}) x(\mathcal{N},t)
\right]/ \sum_{\mathcal{N}=1}^{N} P(\mathcal{N}) \mathcal{N} $ is
the average center of mass position. Here, $P(\mathcal{N})$ is the
probability of having a tube with $\mathcal{N}$ atoms. Assuming  an
initial Thomas-Fermi distribution of the 3D system,
$P(\mathcal{N})\approx 2/3 ({N}^2 \mathcal{N})^{-(1/3)}$
\cite{Paredes}.

When there is at most one atom per site, atoms in the Mott state
have been shown to be responsible for the overdamped dynamics
\cite{Ana2005,Rigol3}. Here, the two-layer model predicts that a
Mott state is created in the lowest layer for $V_o \gtrsim 3 E_R$,
Fig.1, and therefore atoms in this layer are almost frozen. However,
atoms in the second layer are not necessarily localized, and their
dynamics can be underdamped. Because there are always more atoms in
the lowest layer than in the upper one, the overall dynamics is
overdamped, and this explains the large damping observed in the
experiment for $V_o \gtrsim 3 E_R$. This is also in agreement with
the qualitative explanation of the damping given in
Ref.\cite{Soferle}.

Strictly speaking, the fluctuations in the two layers do not overlap
during the dynamics only for $V_o> 7 E_R$, thus we expect the model
to be valid for the deepest lattice depths only. Nevertheless, Fig.4
shows that the theoretical calculations are within the experimental
error bars already for $V_o \gtrsim 4$. Figure 4 also shows that the
damping rates for the central tube and the average are very similar,
and this is because tubes with about $N=80$ atoms have the largest
weight. The average shows a larger damping because it takes into
account contributions from tubes that do not have extra atoms in the
second layer. In the inset  we also compare the experimental center
of mass position of the atomic cloud after $90 ms$ with the model
solutions. The agreement between experiment and theory is consistent
with the one found for the damping rate.

{\it{Summary}} We have developed a simple model that generalizes the
Bose-Fermi mapping to  regimes  where  the filling factor   is
larger than one. The model is relevant for 1D trapped gases where
the co-existence of  Mott-insulating regions with different
occupation numbers is permitted. We presented the necessary
conditions for the model to be valid and showed the uselfuness and
limitations of the method by comparing its predictions for some
physical observables with numerical Monte-Carlo simulations. Very
good agreement between the model and the numerical solutions was
found in the parameter regime where the model is valid. Finally we
used the PF model to study the overdamped dynamics of the center of
mass after a sudden displacement of the trapping potential, and
found good agreement with recent experiments. In particular, the
overdamped motion was linked to the presence of a Mott state in the
lowest atomic layer.\\

The authors thank Trey Porto and Jamie Williams for discussions,
and Nikolay V. Prokof'ev for discussions and for providing the
Monte-Carlo code. This
research was supported in part by ARDA/NSA.

\end{document}